# Dynamical regimes of vortex flow in type-II superconductors with parallel twin boundaries

Harshwardhan Chaturvedi[1], Nathan Galliher[1,2], Ulrich Dobramysl[3],
Michel Pleimling[1,4], and Uwe C. Täuber[1]

[1] Department of Physics & Center for Soft Matter and Biological Physics (MC 0435), Virginia Tech, 850 West Campus Drive, Blacksburg, Virginia 24061, USA

[2] Department of Physics and Astronomy, University of North Carolina, 120 East Cameron Ave, Chapel Hill, NC 27599, USA

[3] Wellcome Trust / CRUK Gurdon Institute, University of Cambridge, Tennis Court Rd, Cambridge CB2 1QN, United Kingdom

[4] Academy of Integrated Science (MC 0405), Virginia Tech, 300 Turner Street NW, Blacksburg, Virginia 24061, USA

**Abstract** We explore the dynamics of driven magnetic flux lines in disordered type-II superconductors in the presence of twin boundaries oriented parallel to the direction of the applied magnetic field, using a three-dimensional elastic line model simulated with Langevin molecular dynamics. The lines are driven perpendicular to the planes to model the effect of an electric current applied parallel to the planes and perpendicular to the magnetic field. A study of the long-time non-equilibrium steady states for varying sample orientation and thickness reveals a rich collection of dynamical regimes spanning the depinning crossover region that separates the pinned and moving-lattice states of vortex matter. We observe the emergence of a preferred direction for the ordering of the Abrikosov lattice in the free-flowing vortex regime due to asymmetric pinning by the planar defects. We have performed novel direct measurements of flux line excitations such as half-loops and double kinks to aid the characterization of the topologically rich flux flow profile.

## 1 Introduction

Point, columnar, and planar quenched disorder serve as natural pinning centers for magnetic-flux lines in type-II superconductors [1]. Effective pinning action through material defects may be used to curb flux flow due to external electric current, thereby mitigating the associated Ohmic loss and leading to a significant decrease in sample resistivity [2–5]. Planar defects are commonly found in the form of twin boundaries in high-$T_c$ cuprates such as (doped) $YBa_2Cu_3O_{7-x}$ (YBCO) and $La_2CuO_{4+\delta}$. Twin boundaries are formed in these materials as they undergo a tetragonal to orthorhombic structural phase transition during the oxidative cooling phase of synthesis [6, 7].

Twin boundaries tend to occur naturally as a mosaic of twins from one of two orthogonal families [8, 9]. Samples containing a single family of twin planes are fabricated artificially [3–5]. The work in this paper pertains to the latter. In the case of a single family of twin planes, the pinning effect of twin boundaries on flux lines is highly anisotropic [4, 10–18], *i.e.*, it strongly depends on the angle between the magnetic field and the twin planes (that are both oriented along the crystallographic *c* axis). Early experiments exploring this variation of pinning strength with field orientation yielded contradictory results [2, 19], with later experiments [10] conclusively showing that pinning is strongest when the field is parallel to the twin planes and the current is flowing in the *ab* plane parallel to the twins thus exerting a Lorentz force on the flux lines perpendicular to the planar defects, confirming the results first presented by Kwok and collaborators [2, 20]. Experiments on flux-boundary pinning can be broadly classified into two types – those that measure electrical transport properties of the system such as resistivity and critical depinning current, and those where the flux lines are directly imaged via techniques like small angle neutron scattering [21] and scanning tunneling microscopy [22].

In transport experiments, planar defects are seen acting as strong pinning centers by their influence on the linear resistivity of a sample near the melting point of the Abrikosov lattice into a flux liquid. The monotonic increase of resistivity with temperature observed in the absence of disorder is interrupted by a drop near the lattice melting transition in the presence of material defects due to the pinning of vortices by the disorder. This drop is more pronounced for spatially correlated disorder such as columnar or planar defects owing to their superior pinning properties [2, 23]. Experiments measuring the critical depinning current density $J_c$ in systems with planar defects also confirm the strong flux-boundary pinning hypothesis, with a sharp maximum in $J_c$ observed as a function of temperature just below the melting point of the Abrikosov lattice in a phenomenon known as the peak effect [24]. Real-time imaging experiments of flux lines driven perpendicular to a single family of twin planes also show strong pinning of magnetic vortices at the twin boundaries [25–27]. However, relatively recent experiments utilizing scanning superconducting quantum interference device microscopy to probe vortex motion near twin boundaries in pnictide superconductors show that vortices avoid pinning to twin boundaries in these materials owing to enhanced superfluid density near the boundaries. They instead prefer to move parallel to them. This flux-boundary repulsion is offered as a possible explanation for the enhanced critical currents observed in twinned superconductors [28].

Numerical studies of vortex behavior in the presence of planar defects range from solving the full time-dependent Ginzburg-Landau equations [29–32] to more approximate descriptions [33–37] of vortices in two-dimensional thin-film and three-dimensional bulk samples as structureless point- or string-like objects that are studied with either Monte Carlo simulations or Langevin dynamics methods. The experimentally detected anisotropy of pinning and transport has been observed in numerical simulations of twinned superconductors [38, 39] with thermal fluctuations being enhanced and vortex motion facilitated within defect planes. Reichhardt *et al.* identified three phases of flux flow in London-Langevin studies of driven vortices subject to planar pinning, *viz.* guided plastic flow at low drives characterized by partially-ordered vortices, highly disordered plastic flow at intermediate drives, and elastic flow at high drives with the vortices reordering into a lattice in this phase [34]. This is in agreement with earlier results of Crabtree *et al.* [40] which were obtained by solving the time-dependent Ginzburg-Landau equations on a discrete grid. Recent analytical works argue that for randomly placed parallel planar defects, the flux line lattice displays a novel planar glass phase with an exponential decay of long-range translational order as opposed to the algebraic decay seen in the case of the Bragg glass obtained in the absence of planar defects [41–43]. In their analytical studies of the low-temperature dynamics of magnetic flux lines in type-II superconductors, Marchetti and Vinokur find that flux lines driven transverse to a family of parallel twin planes do so in a manner analogous to the motion of one-dimensional charge carriers in a disordered semiconductor induced by an electric field. They discuss various linear and non-linear transport mechanisms for vortex motion that are associated with different flux line excitations in the system [44, 45].

In this present work, we use an elastic line description of vortices in a three-dimensional sample modeled to mimic the behavior of flux vortices in the mixed phase of YBCO. The elastic lines are mutually repulsive and are subject to a horizontal drive representing the Lorentz force exerted by an external current. The sample contains two planar defects perpendicular to the drive direction as well as many ran-domly distributed point-like pinning sites that represent point disorder such as those produced by oxygen vacancies (Fig. 1). The dynamics of this model is simulated by numerically solving overdamped Langevin equations that account for the fast degrees of freedom in the system as stochastic forcing that is subject to certain physical constraints. This particular implementation of the elastic line model was previously used by Dobramysl *et al.* [46] to study relaxation and aging phenomena of flux lines in the presence of point-like and columnar disorder. Since then, it has been employed to investigate relaxation dynamics of vortex lines following magnetic field, temperature and drive quenches [47–49], as well as the pinning time statistics for flux lines in disordered environments [50]. We have extended this work to here address the dynamics of vortices driven parallel to the $x$ axis, and perpendicular to two parallel planar defects that are placed either a short distance (16 pinning center radii $b_0$) apart or a large distance (160$b_0$) apart. The system is periodic in the $x$ direction and therefore the planar defect pair configuration employed here is comparable to a long YBCO sample containing evenly spaced pairs of parallel twin boundaries. We observe the long-time steady-state behavior of this system of flux lines for two sample orientations (aspect ratios) and several sample thicknesses $L$, *i.e.,* system extensions along the magnetic field or $z$ direction. These observations involve measuring several physical attributes and occurrence statistics for different flux line excitations as well as static and dynamic visualizations of the system under a range of conditions and from a number of (both two- and three-dimensional) perspectives.

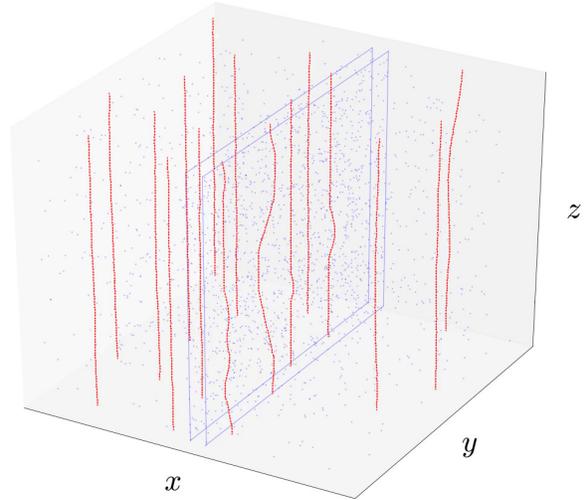

**Fig. 1** Simulation snapshot of flux lines (red) driven along the $x$ axis in the presence of two planar defects (blue) oriented perpendicular to the direction of drive and many randomly positioned point defects (blue).

The characterization of the depinning process, by which magnetic vortices subject to planar pinning transition from the pinned to the moving lattice state, has been greatly enhanced by direct measurements of the unique vortex excitations that emerge from planar defect-induced elastic deformations of these vortices. These measurements are made possible by the full three-dimensional specification of our simulated model coupled with the structural simplicity of the infinitesimally thin elastic lines that represent the vortices. The steady-state results pertaining to the depinning region reveal a rich assortment of drive regimes that culminate in the dynamical freezing of the vortices into a moving hexagonal lattice. Upon increasing the driving force, distinct crossover regimes are encountered between the fully pinned glassy phase and the freely flowing ordered lattice, depending on the orientation of the system. In these intermediate current regimes, the vortices remain partially pinned to the planar defects; flux transport in these regimes is mediated through vortex half-loop, single-kink, and double-kink excitations. The quantitative analysis of vortex excitations complement these results by providing us with insight into the types of structures that facilitate the realization of the different depinning regimes.

The organization of this paper is as follows. Section 2 explains the various terms in the Hamiltonian for our elastic line model, describes the Langevin Molecular Dynamics algorithm we employ to implement its stochastic dynamics, and specifies the material parameters we use for the implementation. This section also covers definitions of the seven observable quantities we measure directly and the simulation protocol we use to evolve the system to the steady state. We discuss the relevant results in Section 3. These include the different regimes of flux creep and flow, and the preferred arrangement of vortices induced by strong anisotropic pinning. We conclude the paper by summarizing our results in Section 4.

## 2 Elastic Line Model and Simulation Protocol

### 2.1 Model Hamiltonian

We model flux lines as mutually repulsive elastic lines [51, 52] in the extreme London limit, *i.e.*, when the London penetration depth is much larger than the coherence length. The Hamiltonian of the system is a sum of four terms, *viz.* the elastic line tension energy, the attractive potential due to pinning sites, the repulsive pair interactions between vortex line elements, and the work done by the external electric current:

$$
\begin{aligned}
H[\mathbf{r_i}] = \sum_{i=1}^{N} \int_0^L dz &\left[ \frac{\tilde{\varepsilon}_1}{2} \left| \frac{d\mathbf{r_i}(z)}{dz} \right|^2 + U_D(\mathbf{r_i}(z)) \right. \\
&\left. + \frac{1}{2} \sum_{j \neq i}^{N} V(|\mathbf{r_i}(z) - \mathbf{r_j}(z)|) - \mathbf{F_d} \cdot \mathbf{r_i}(z) \right].
\end{aligned} \tag{1}
$$

$\mathbf{r}_i(z)$ represents the position vector in the $xy$ plane of the line element of the $i$th flux line (one of $N$), at height $z$. The elastic line stiffness or local tilt modulus is given by $\tilde{\varepsilon}_1 \approx \Gamma^{-2} \varepsilon_0 \ln(\lambda_{ab}/\xi_{ab})$ where $\Gamma^{-2} = M_{ab}/M_c$ is the effective mass ratio or anisotropy parameter. $\lambda_{ab}$ is the London penetration depth and $\xi_{ab}$ is the coherence length, in the $ab$ crystallographic plane. The in-plane repulsive interaction between any two flux lines is given by $V(r) = 2\varepsilon_0 K_0(r/\lambda_{ab})$, where $K_0$ denotes the zeroth-order modified Bessel function. It effectively serves as a logarithmic repulsion that is exponentially screened at the scale $\lambda_{ab}$. The pinning sites are modeled as smooth potential wells, given by

$$
U_D(\mathbf{r}, z) = -\sum_{\alpha=1}^{N_D} \frac{b_0}{2} p \left[ 1 - \tanh\left( 5\frac{|\mathbf{r} - \mathbf{r_\alpha}| - b_0}{b_0} \right) \right] \\
\times \delta(z - z_\alpha), \tag{2}
$$

where $N_D$ is the number of pinning sites, $p \geq 0$ is the pinning potential strength, $b_0$ is the width of the potential well, while $\mathbf{r}_\alpha$ and $z_\alpha$ respectively represent the in-plane and vertical positions of pinning site $\alpha$. Each potential well is smooth at its boundary, and drops steeply to a flat minimum $-b_0 p/2$ in its bottom.

We employ periodic boundary conditions in the $x$ and $y$ directions and free boundary conditions in the $z$ direction. We configure the horizontal $xy$ plane of the system in one of two equal-area, orthogonal orientations: (A) $16/\sqrt{3}\lambda_{ab} \times 8\lambda_{ab}$ or (B) $8\lambda_{ab} \times 16/\sqrt{3}\lambda_{ab}$. Orientation B is a 90° rotation of orientation A about the $z$ axis. The two orientations produce markedly different flux flow profiles that are discussed in the results section below. The particular ratio of horizontal boundary lengths is necessary to ensure that the flux lines can equilibrate to a periodic hexagonal Abrikosov lattice. The Lorentz force exerted on the flux lines by an external current $\mathbf{j}$ is modeled in the system as a tunable, spatially uniform drive $F_d = |\mathbf{j} \times \phi_0 \mathbf{B}/B|$ in the $x$ direction. All lengths are measured in units of $b_0$ while energies are measured in units of $\varepsilon_0 b_0$, where $\varepsilon_0 = (\phi_0/4\pi\lambda_{ab})^2$ is the elastic line energy per unit length, and $\phi_0 = hc/2e$ is the magnetic flux quantum.

In each simulation run, we set the orientation and sample thickness of the system (in the $z$ direction) $L$ to the desired values and initialize it with two planar defects oriented perpendicular to the direction of the drive ($x$ direction). Each planar defect consists of columns of point defects extending along the entire height of the system. These columns are stacked side by side along the $y$ direction, and consecutive defects are separated by a distance of $2b_0$. We set up our pair of defect planes in one of two configurations – either close together, *i.e.*, where the planes are separated by $16b_0$ (∼ 5% of the system length in the $x$ direction), or far apart with a separation of $160b_0$ (∼ 50% of the system length in the $x$ direction). Besides the two defect planes, isolated point defects are randomly distributed throughout the system to maintain a concentration of 1116 defects per plane. The random point defects provide the effective viscosity experienced by moving flux lines in a real physical system.

## 2.2 Langevin Molecular Dynamics

We simulate the dynamics of the model by discretizing the system along the direction of the external magnetic field ($z$ direction) into layers. Consecutive layers are separated by $c_0$, i.e., one crystal unit cell size along the crystallographic $c$ direction [52, 53]. Consequently, each elastic line consists of elastically coupled points, with each discrete element residing in a unique layer. The pinning sites (2) are also confined to these layers. The interactions between these discrete elements are encapsulated in the properly discretized version of the Hamiltonian (1) that we use to obtain coupled overdamped Langevin equations which we then solve numerically:

$$\eta \frac{\partial \mathbf{r}_i(t, z)}{\partial t} = -\frac{\delta H[\mathbf{r}_i(t,z)]}{\delta \mathbf{r}_i(t,z)} + \mathbf{f}_i(t,z). \quad (3)$$

Here, $\eta = \phi_0^2/2\pi\rho_n c^2 \xi_{ab}^2$ is the Bardeen-Stephen viscous drag parameter, where $\rho_n$ represents the normal-state resistivity of YBCO near $T_c$ [1, 54]. We model the fast, microscopic degrees of freedom of the surrounding medium by means of thermal stochastic forcing as uncorrelated Gaussian white noise $\mathbf{f}_{i,z}(t)$ with vanishing mean $\langle \mathbf{f}_{i,z}(t)\rangle = 0$. Furthermore, these stochastic forces obey the Einstein relation

$$\langle \mathbf{f}_{i,z}(t) \cdot \mathbf{f}_{j,z'}(s)\rangle = 4\eta k_B T\, \delta_{ij}\delta_{zz'}\delta(t-s),$$

which ensures that the system relaxes to thermal equilibrium with a canonical probability distribution $P[\mathbf{r}_{i,z}] \propto e^{-H[\mathbf{r}_{i,z}]/k_B T}$ in the absence of any external current.

## 2.3 Model Parameters

We have selected our model parameters to closely match the material properties of the high-$T_c$ type-II superconductor YBCO. The pinning center radius is set to $b_0 = 35\text{Å}$. The inter-layer spacing in the crystallographic $c$ direction is set to $c_0 = b_0$. The in-plane London penetration depth and superconducting coherence length are chosen to be $\lambda_{ab} = 34b_0 \approx 1200\text{Å}$ and $\xi_{ab} = 0.3b_0 \approx 10.5\text{Å}$, respectively, in order to represent YBCO, which has a high effective mass anisotropy ratio $\Gamma^{-2} = 1/25$. The line energy per unit length is $\varepsilon_0 \approx 1.92 \cdot 10^{-6}\text{erg/cm}$. This effectively renders the vortex line tension energy scale to be $\tilde{\varepsilon}_1/\varepsilon_0 \approx 0.189$. The pinning potential well depth is taken as $p/\varepsilon_0 = 0.05$. The temperature in our simulations is set to 10 K ($k_B T/\varepsilon_0 b_0 = 0.002$ in our simulation units). The Bardeen–Stephen viscous drag coefficient $\eta = \phi_0^2/2\pi\rho_n c^2 \xi_{ab}^2 \approx 10^{-10}$ erg·s/cm$^2$ is set to one, where $\rho_n \approx 500\ \mu\Omega$m is the normal-state resistivity of YBCO near $T_c$ [55]. This results in the simulation time step being defined by the fundamental temporal unit $t_0 = \eta b_0/\varepsilon_0 \approx 18$ ps; simulation times are measured in units of $t_0$.

## 2.4 Measured Quantities

Our understanding of the system is primarily developed by examining certain relevant physical observables. One such observable is the *local hexatic order parameter*

$$h = \left\langle \frac{1}{m_i}\sum_{j=1}^{m_i}\cos(6\theta_{ij})\right\rangle_{i,k}, \quad (4)$$

a measure of local sixfold orientational order in the system. An $h$ value of 0 indicates a high degree of orientational disorder in the system while $h \approx 1$ signifies a highly ordered hexagonal crystal-like structure. In order to compute $h$, we first compute the mean $xy$ positions of the flux lines in the system. This gives us a two-dimensional representation of the flux lines as points in a plane. For each line in this representation, we identify its nearest neighbors, i.e., other lines that are within a cutoff distance $4/\sqrt{3}\lambda_{ab}$ of itself. The cutoff distance is derived from the distance that separates neighboring lines in an ideal hexagonal Abrikosov lattice. Say line $i$ has $m_i$ nearest neighbors; we then compute the angle $\theta_{ij}$ that each neighbor $j$ makes with the neighbor closest to it in the clockwise direction, and subsequently calculate $\cos(6\theta_{ij})$ for each $j$ and find the mean of these cosines. This value is the measure of hexatic order for line $i$ with respect to its nearest neighbors. In the final expression for $h$ stated above, $\langle\ldots\rangle_{i,k}$ represents an average over all $N$ vortex lines $i$ and different realizations $k$ of the disorder configurations and noise histories.

Another quantity of interest is the mean *radius of gyration*,

$$r_g = \sqrt{\langle(\mathbf{r}_i(z) - \langle\mathbf{r}_i\rangle_z)^2\rangle}, \quad (5)$$

i.e., the standard deviation of the lateral positions $\mathbf{r}_i(z)$ of the points constituting the $i$th flux line, averaged over all the lines. $r_g$ is a measure of overall roughness of the lines in the system. Here, $\langle\ldots\rangle_z$ represents an average over all layers $z$, while $\langle\ldots\rangle$ represents an average over layers $z$ of line $i$ as well as an average over all lines and different realizations of the disorder and the noise.

Our third observable is the *mean vortex velocity*

$$\mathbf{v} = \left\langle\frac{d}{dt}\mathbf{r}_i(z)\right\rangle. \quad (6)$$

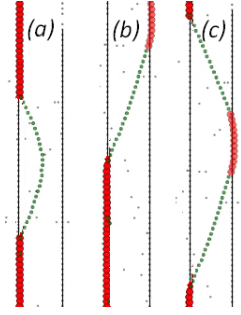

**Fig. 2** Simulation snapshots of the system projected onto the *xz* plane for a side view, showing flux lines forming three different excitations: (a) half-loop, (b) single kink, and (c) double kink. The green dotted lines are flux lines, the gray dots represent point pins, and the black vertical lines planar defects. The red dotted sections are those portions of the flux lines that are trapped at planar defects. The drive $F_d$ is oriented along the positive *x* (right) direction.

The fourth quantity we measure is the *fraction of pinned line elements*

$$f_p = \langle n(r < b_0)/n_{\text{total}} \rangle_k \ . \tag{7}$$

Here, $n(r < b_0)$ denotes the number of line elements located at a distance $r$ less than one pinning center radius $b_0$ from a pinning site. $n_{\text{total}}$ is the total number of line elements in the system. Thus, $f_p$ is the fraction of line elements in the system that are located within distance $b_0$ of an attractive defect site. Here, $\langle \dots \rangle_k$ represents an average over different realizations $k$ of the disorder and the noise. We remark that we initially obtained the separate contributions to the pinning fraction that respectively originated from planar and point defects. We found that the overall pinning is dominated mostly by the planar defects. For conciseness, we have only reported results pertaining to the overall pinning fraction in this paper.

We also measure the numbers of different flux line excitations that appear in the system, *viz. half-loops*, *single kinks*, and *double kinks* (Fig. 2). A flux line forms a half-loop (Fig. 2a) when it becomes partially depinned from a defect plane and the separation between the depinned portion and the plane is smaller than the inter-planar distance. A single kink (Fig. 2b) appears when part of a line is trapped in one defect plane while an adjacent section is trapped in the neighboring plane. A double kink (Fig. 2c) is similar to a half-loop but with a larger separation between the depinned portion and the remainder of the flux line that results in the outermost portion of the half-loop being pinned to the next defect plane; this can also be viewed as a specific combination of two single kinks and is accounted for as such in our measurements. In each simulation run, we record the total number of each type of vortex excitation appearing in the system. The excitation numbers depend on the total number of flux lines $N$ in the system. We have used $N = 16$ vortices throughout this study.

### 2.5 Simulation Protocol

We obtain steady-state results for our system via the following procedure. We randomly place $N = 16$ straight flux lines in the system and immediately subject them to an effective temperature of $0.002\varepsilon_0 b_0/k_B$ and the desired drive strength $F_d$. The lines are allowed to relax in this constant temperature-drive bath for an initial relaxation time of $100,000t_0$. At this point, we start measuring the various observables in the system every 100 time steps, a duration larger than the correlation times in the system that range from $20t_0$ to $45t_0$ depending on the strength of the applied driving force. We perform 1000 such measurements and under the ergodic assumption, record their average for each observable. We simulate 10 independent realizations in this manner and perform an ensemble average over these realizations. Between the time averaging and ensemble averaging, we thus average each data point over $10,000$ independent values.

## 3 Results

We have performed a detailed study of the steady-state behavior of vortex matter subject to two parallel extended planar defects oriented perpendicular to the drive for two orientations of the system and several sample thicknesses (ranging from $L = 50b_0$ to $250b_0$). These steady-state curves reveal drive regimes not observed in our previous studies [46, 48] with point-like and columnar defects, and underscore the richer kinetics accessible with this defect geometry.

### 3.1 Depinning Drive Regimes and Preferred Ordering

We begin with results for the case where the planar defects are placed closely together. In both orientations, A and B, we see a *pinned regime* at the beginning of, and a *moving-lattice regime* at the end of the driving-force range under consideration. For orientation A, where the system is more extended along the $x$ than the $y$ direction, we find three intermediate regimes (*liquid*, *partially-ordered*, and *smectic*) that span the depinning crossover region connecting the two extremal regimes (Fig. 3) [56], while for orientation B (simulation domain longer in the $y$ direction), there exists only one intermediate *liquid / smectic* regime. In this section we discuss the mechanisms underlying the development of the differing crossover flow profiles in the two orientations. In our simulations with rather few interacting vortex lines, the boundaries between these distinct drive regimes are not well defined; the crossover regions separating them have non-zero widths and their sizes are not uniform (Fig. 3). These steady-state results are supplemented by simulation snapshots of the system under different drive conditions, which provide visual evidence for the defining structural configurations that the flux lines assume in different regimes.

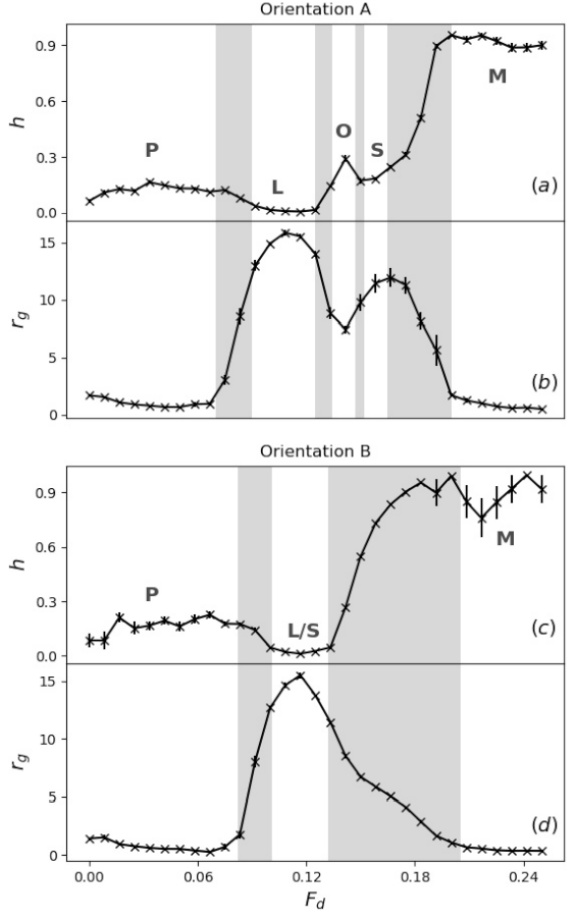

**Fig. 3** Steady-state (a, c) local hexatic order parameter $h$, and (b, d) radius of gyration $r_g$ (units of $b_0$) for interacting flux lines in the presence of two closely placed planar defects, in a sample of thickness $L = 250b_0$. The top two figures (a, b) display results for the system in orientation A (simulation domain longer in the $x$ direction along the drive) and the bottom two (c, d) do so for orientation B (simulation domain more extended along the $y$ direction). Vertical gray bars are used to indicate the crossover regions that separate consecutive drive regimes. The regimes are labeled with the acronyms *P*: pinned, *L*: liquid, *O*: partially-ordered, *S*: smectic, *M*: moving lattice, and *L/S*: liquid/smectic. Note that these data are replicated in Fig. 4. Here and in the following figures, only error bars larger than the symbol sizes are shown.

### 3.1.1 The Pinned Regime

At low driving currents, the first dynamical steady-state region encountered in the system is the pinned regime. Observed in both orientations, the pinned regime is characterized by zero mean velocity $v$ (Fig. 4a/e) and a sizable fraction of pinned line elements $f_p$ (Fig. 4b/f). The degree of local hexatic order $h$ is relatively low ($\sim 0.2$ for $L = 250b_0$) as seen in Fig. 3a/c. In this regime, a proportion of the flux lines are trapped in the first planar defect (the one with the lower $x$ coordinate) while the remainder are held stationary at a fixed distance behind them by the long-range inter-vortex repulsions. This is visible in Fig. 5a1 and 5a2. Although the snapshots in these figures are taken for the system in orientation A, they are qualitatively faithful representations of vortex behavior in the pinned regime for orientation B as well. Due to the vertically correlated configuration of the planar defects (along the magnetic field or $z$ direction), the flux lines trapped within them are nearly perfectly straight and therefore display a low radius of gyration (Fig. 4d/h). Lateral fluctuations for the unpinned vortices are suppressed as well on account of both their intrinsic elastic line tension and the repulsive caging induced by the lines trapped in the defect planes positioned in front of them. This results in the low overall gyration radius we observe in the steady state in the pinned regime. Note that the fraction of pinned line elements $f_p$ actually grows monotonically with drive (Fig. 4b/f): higher driving forces induce larger pinning fractions since an increase in drive shrinks the distance between the free-standing caged vortices and those trapped in the first defect plane, thereby increasing the likelihood of a free-standing vortex being pulled into that extended defect. Correspondingly, the influence of sample thickness $L$ on all the measured quantities is negligible, with the steady-state curves for different $L$ appearing identical within our statistical errors, since the flux lines are virtually motionless barring thermal fluctuations in the pinned regime. The influence of the flux line length only becomes appreciable once the vortices start moving, as described below.

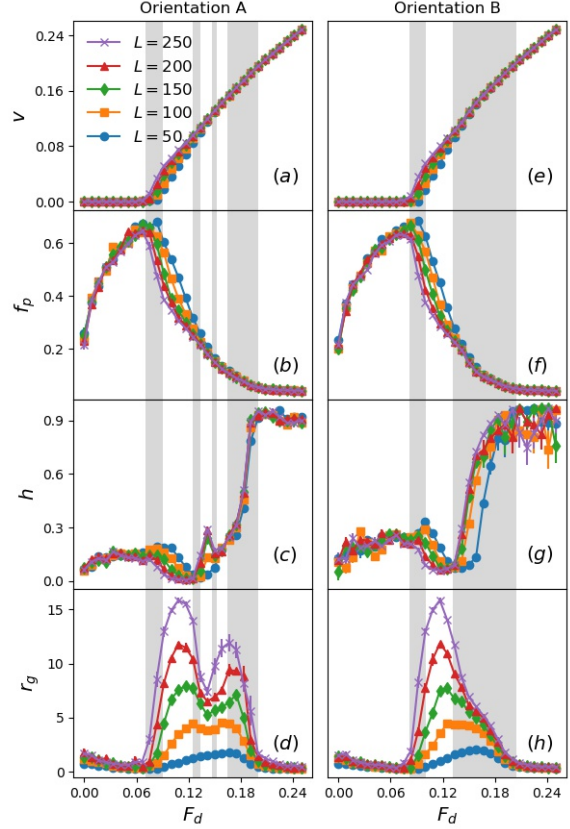

**Fig. 4** Steady-state (a, e) mean vortex velocity $v$ (units of $b_0/t_0$); (b, f) fraction of pinned line elements $f_p$; (c, g) local hexatic order parameter $h$; and (d, h) radius of gyration $r_g$ (units of $b_0$) as a function of drive $F_d$ (units of $\varepsilon_0$) for interacting flux lines in the presence of two closely placed planar defects, in samples of varying sample thickness $L$ (units of $b_0$). The figures on the left (a, b, c, d) display results for the system in orientation A and those on the right (e, f, g, h) do so for orientation B. Vertical gray bars are used to indicate the crossover regions that separate consecutive drive regimes for $L = 250b_0$.

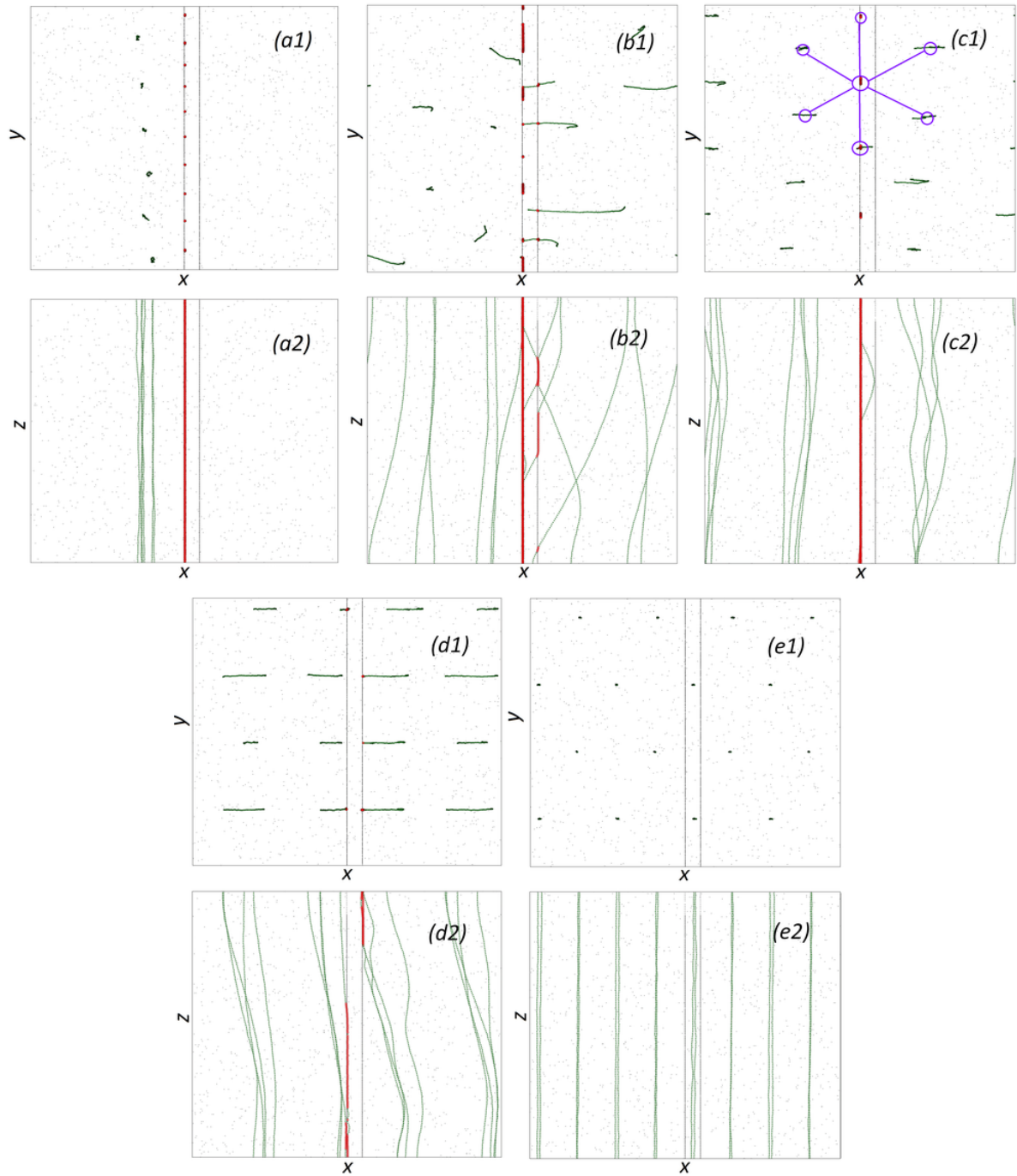

**Fig. 5** Simulation snapshots of a system with 16 interacting flux lines in a sample with orientation A and thickness $L = 150b_0$, in the presence of two closely placed planar defects, projected onto the (1) $xy$ plane for a top view, and (2) $xz$ plane for a side view, in respectively the (a) pinned ($F_d = 0.06\varepsilon_0$), (b) liquid ($F_d = 0.11\varepsilon_0$), (c) partially-ordered ($F_d = 0.14\varepsilon_0$), (d) smectic ($F_d = 0.16\varepsilon_0$), and (e) moving-lattice ($F_d = 0.22\varepsilon_0$) drive regimes. The green dotted lines represent the vortices, the black vertical lines indicate planar defects. The red dotted sections mark those portions of the flux lines that are trapped by planar defects. The purple circles and lines in (c1) indicate a central vortex and its nearest neighbors forming a hexagon-like structure. The drive $F_d$ is oriented in the positive $x$ (right) direction. The system boundary lengths in the $x$ and $y$ directions are $314b_0$ and $272b_0$, respectively. The full videos from which these snapshots have been taken can be viewed at https://figshare.com/s/8a1e4bf34f463f988ebd.

### 3.1.2 The Liquid Regime

Increasing the drive further, we exit the pinned regime and begin entering the liquid regime for orientation A and the liquid/smectic regime for orientation B. We explain the differentiating factors between these regimes in the next subsection on the partially-ordered and smectic regimes. During the crossover from the pinned to the liquid(/smectic) regime, the drive is sufficiently strong for the vortices to detach from the defect planes. In the course of depinning, the flux lines suffer large distortions (Fig. 5b1 and Fig. 5b2) that result in a sharp increase of their mean gyration radius (Fig. 3b/d). The degree of orientational disorder is maximized as the local hexatic order parameter $h$ approaches zero (Fig. 3a/c) and remains suppressed for the extent of the regime. The incipient vortex motion naturally results in the mean vortex velocity assuming non-zero values (Fig. 4a/e). Both gyration radius and mean vortex velocity increase monotonically with drive, while the pinning fraction diminishes (Fig. 4b/f) as the driving force is increased to enter deeper into the depinning region.

The crossover from the pinned into the liquid regime occurs at lower drive values for thicker samples (greater $L$) as evidenced in all observables shown in Fig. 4. This is because longer vortex lines comprise a larger number of segments along their trajectory which can potentially be set free from the defect plane holding them, by the applied drive with assistance from thermal fluctuations. This increases the probability of neighboring line elements to break free as they are elastically coupled to the first detached element, inducing a cascading effect whereby the entire line is pulled free from the defect plane. For any given drive in this regime, longer flux lines display a larger gyration radius (Fig. 4d/h), since they are capable of incorporating larger distortions as they are pulled away from the defect planes at different locations along their contour. The opposite trend holds true for the pinning fraction $f_p$ (Fig. 4b/f), with shorter lines being more likely to remain trapped by planar defects. The propensity to be partially depinned also results in longer lines moving faster on average (Fig. 4a/e) as their motion is less impeded by the disorder.

### 3.1.3 The Partially-Ordered and Smectic Regimes

As the drive is increased beyond the liquid(/smectic) regime, we see differences develop in the steady-state behavior for different system orientations. For orientation A, the system begins to develop local hexatic order (Fig. 3a), which marks the onset of the partially-ordered regime, the second of the three intermediate regimes in the depinning region. In our simulations, the sixteen flux lines arrange themselves into eight horizontal pairs flowing along the positive $x$ direction as seen in Fig. 5c1. The pairs are approximately equally spaced in the $y$ direction. In this regime, we observe the formation of hexagon-like structures in the vicinity of the planar defects. Fig. 5c1 illustrates that in the $xy$ plane, each approximate hexagon consists of a central vortex trapped at a planar defect surrounded by six nearest-neighbor flux lines. Two of the adjacent vortices are located in the row above the central one, while two are in the row below. The fifth and sixth nearest neighbors are symmetrically situated in the second row above and below the central vortex, respectively. Each pair of consecutive neighbors subtends an angle of $\approx 60°$ at the central vortex. The organization of the lines into this hexagonal lattice-like configuration results in each line experiencing enhanced repulsive caging by its neighbors giving rise to comparatively straighter flux lines (Fig. 5c2). This is evident in the reduced gyration radius $r_g$ (Fig. 3b) in this region. The stronger external forcing propels the flux lines faster through the defect planes, as seen in the rising drive-velocity (or current-voltage) curves (Fig. 4a) for all sample thicknesses $L$. The pinning fraction continues to decline with drive, albeit at a slower rate than in the previous regime (Fig. 4b).

The partially-ordered regime seen for orientation A exists in a rather small driving force interval. As the drive is increased further, the modest gains in local hexatic order $h$ in the partially-ordered regime are rapidly lost again (Fig. 4c), as the more disordered smectic regime is reached. The eight horizontal vortex pairs of the partially-ordered regime give way to four distinct horizontal flux flow channels that are akin to dynamic smectic ordering. Each channel consists of four flux lines in our simulations; the channels are equally spaced along the $y$ axis. This geometric arrangement is characterized by a larger typical inter-vortex distance. This essentially gives the flux lines more wiggle room and subjects them to weaker repulsive caging by neighboring lines as compared to the partially-ordered regime. This is evident visually from the extended nature of the flux lines along the $x$ axis as seen in Fig. 5d1, and quantitatively from the enhanced radius of gyration $r_g$ (Fig. 4d). Mean vortex veloc-ity and pinning fraction continue to monotonically in- and decrease, respectively, within the entire smectic region, with the four-row smectic eventually crystallizing into a four-row moving lattice.

We explain this rather peculiar sequence of orientational-order transitions for system orientation A, *i.e.*, disorder-to-order (liquid to partially-ordered), followed by order-to-disorder (partially-ordered to smectic), followed again by disorder-to-order (smectic to moving lattice), via the following mechanism: The partially-ordered regime consists of eight rows and four columns of flux lines, with each row containing two vortices and each column comprising four (Fig. 5c1). The geometry of this configuration is similar to that of the ultimate depinned moving lattice, which consists of the flux lines sorted into four rows and eight columns (Fig. 5e1). The partially-ordered eight-row pseudo-lattice is actually an approximation of the highly ordered four-row moving lattice rotated by 90° about the $z$ axis. In the absence of defects, the flux lines in orientation A will naturally form a four-row moving lattice since this configuration is more energetically favorable in this orientation of the system (containing a longer $x$ boundary). However, strong anisotropic pinning in one direction favors a vortex lattice configuration that maximizes the number of line segments being pinned at any given time. This translates to the planar defects along the $y$ direction favoring a vortex lattice with the maximum number of vortices along the $y$ direction (*i.e.* per column), which happens to be the eight-row, four-column rotated lattice. The vortex configuration favored by the anisotropic planar pinning is thus distinct from, and competing against the flux line structure that arises naturally in this system orientation. An approximation of the former (the eight-row partially-ordered configuration) is energetically favorable for a brief drive interval just as the vortex system begins to depin into the liquid regime, while the effective pinning strength is not yet overpowered by the drive. Yet at elevated driving forces, the influence of the planar defects predictably weakens, resulting in the re-shuffling of the vortices into the four-row smectic that is a precursor to the natural ordering of flux lines favored by the system orientation A.

Further evidence for preferred vortex lattice ordering by the planar disorder is found in the simulation results for orientation B. Here, the system boundary length in the $y$ direction is greater than that in the $x$ direction, and consequently, the natural configuration for the vortex lattice is the eight-row, four-column version, which is orthogonal to the expected natural configuration for orientation A. Note that for this orientation, the default vortex lattice configuration in the absence of defects also happens to be the arrangement preferred by the planar defects parallel to the $y$ axis, according to our hypothesis of maximal vortex pinning. Unlike in the case of orientation A, in orientation B there are no competing vortex lattice configurations. This is borne out by the fact that for orientation B, there is no rearrangement of the vortices at higher drives, when the pinning effectiveness of the defects is diminished. The vortices in the liquid state begin forming an eight-row smectic by the end of the *liquid/smectic* regime (similar to that in Fig. 5c1), and continue to monotonically crystallize into a highly ordered (Fig. 4g) eight-row hexagonal lattice. This is supported by the absence of the local maximum in hexatic order $h$ (Fig. 3c) seen for orientation A that signifies the onset and cessation of the partially-ordered regime, as well as the absence of the corresponding local minimum in gyration radius (Fig. 3d).

The partially-ordered regime is not observed in thin samples with $L \leq 50b_0$, and is markedly suppressed for lines shorter than $150b_0$ (Fig. 4c). For shorter flux lines, the system directly transitions from the liquid to the moving-lattice regime, forming a smectic along the way, but completely bypasses the partially-ordered regime. As in the case of orientation B, this claim is supported by the missing first local maximum in the $h$ curve for $L = 50b_0$ (Fig. 4c), and the absence of the local minimum in the $r_g$ curve (Fig. 4d). The partially-ordered regime likely constitutes a metastable state that (for any system orientation) is inaccessible to short and stiff vortices which can be approximated as one-dimensional objects moving in a two-dimensional domain. This is supported by the significant similarities in the steady-state profiles of $h$ and $r_g$ at low $L$ for different system orientations (Fig. 4). Sample thickness and hence vortex line length $L$ plays a crucial role in allowing the system to access this metastable region. An increased line length enhances the stochasticity in the system as represented by the $z$-dependent noise term in the Langevin equations (3), thus enabling it to sufficiently explore the energy landscape and find the metastable partially-ordered regime.

The peculiar flux flow profile observed for orientation A is a finite-size effect – a consequence of the particular choice of system boundary lengths in our small system of 16 vortices. An experimental system would be one to several orders of magnitude larger in both extension (in the $x$ and $y$ directions) and the number of vortices. The vortex lattice orientation in such a system would most likely not depend on the ratio of boundary lengths (as is the case in our simulations), and in the absence of strong anisotropic pinning, several orientations of the vortex lattice would be equally likely. However, the results of our study for orientation A, especially when compared to those for orientation B, strongly suggest that the presence of strong anisotropic pinning, such as that due to parallel planar defects, should break rotational symmetry and prefer flux line arrangments that would maximize the number of vortices encountered by the planar defects during flux flow.

### 3.1.4 The moving-lattice regime

The final drive regime for both orientations (A and B) is the moving-lattice regime. The driving current is sufficiently strong that the pinning to the attractive defects becomes negligible, and the lines are once again almost perfectly straight as in the low-drive pinned regime (Fig. 5e2). The flux flow channels along the $x$ direction that had formed in the smectic (or liquid/smectic) regime persist into the moving-lattice regime. With the destabilizing influence of disorder effectively removed, the flux lines arrange themselves within these channels to form a moving hexagonal Abrikosov lattice (Fig. 5e1), with the local hexatic order $h$ approaching unity (Fig. 4c/g). The moving lattice marks the completion of the dynamic freezing process. The mean velocity of the lines shows a linear dependence on drive strength, indicating that the system has entered an Ohmic regime with linear $I$-$V$ characteristics (Fig. 4a/e). The pinning fraction (Fig. 4b/f) plateaus a little above zero (at $f_p \approx 0.04$) as does the gyration radius (Fig. 4d), as the lines move practically freely through the system without the pinning and roughening caused by the defects. The role of sample thickness or vortex line length becomes negligible, with the $h$, $r_g$, $v$, and $f_p$ curves for different $L$ practically coinciding for the extent of the entire moving-lattice regime (Fig. 4).

## 3.2 Flux Line Excitations

In order to better understand the flux line structures that characterize the distinct regimes observed in our studies of vortex matter subject to planar defects, we performed direct measurements of flux line excitations, *viz.* half-loops, single kinks, and double kinks, which appear in our system due the interactions of the flux lines with the closely placed attractive defect planes (see Section 2.4 for our operational definitions of these vortex excitations).

The two local gyration radius maxima (Fig. 4d) that mark the liquid and smectic drive regimes for orientation A coincide exactly with corresponding peaks in the single-kink (Fig. 2b) steady-state curves (Fig. 6b). In the case of orientation B, the steady-state gyration radius (Fig. 4h) and single-kink population (Fig. 6e) move practically together with non-zero drive, indicating that the two quantities are strongly correlated with regard to their evolution with applied driving force, in either system orientation. For all drive strengths, we notice a positive correlation between the number of single kinks and vortex length (or sample thickness) $L$: longer flux lines afford considerably more possible locations along their contours for kinks to form. This correlation between line length and number of steady-state excitations holds true for the other line structures (half-loops and double kinks) under consideration as well (Fig. 6).

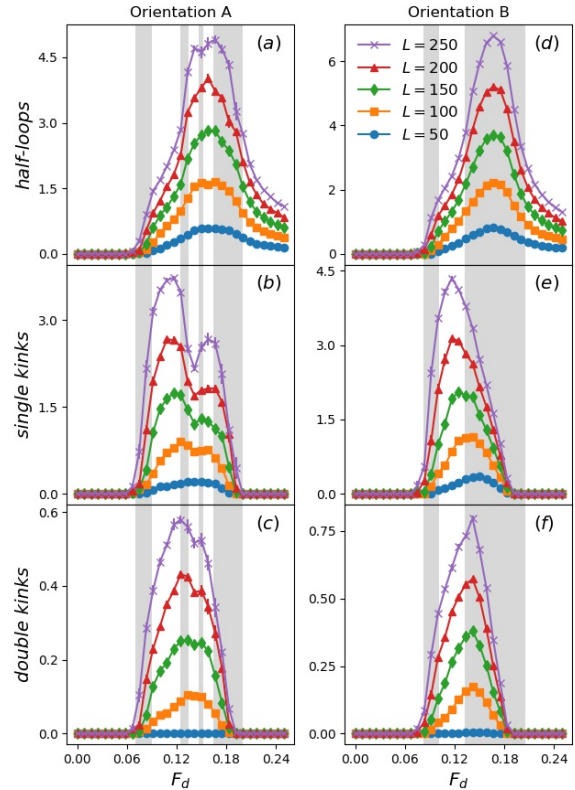

**Fig. 6** Steady-state number of (a, d) half-loops, (b, e) single kinks, and (c, f) double kinks as a function of drive $F_d$ (units of $\varepsilon_0$) for interacting flux lines in the presence of two closely placed planar defects, in samples of varying sample thickness $L$ (units of $b_0$). The figures on the left (a, b, c) display results for the system in orientation A and those on the right (d, e, f) do so for orientation B. Vertical gray bars are used to indicate the crossover regions that separate consecutive drive regimes for $L = 250b_0$.

Vortex half-loops (Fig. 2a) occur in the system with a frequency comparable to that of single kinks. They peak at the end of the smectic regime (Fig. 6a) for orientation A, and within the crossover region separating the liquid/smectic and moving lattice regimes for orientation B. In contrast to single kinks, the formation of half-loops requires the flux lines to be relatively straight. We thus observe a steady increase in the number of half-loops beyond the pinned regime as the lines that start out distorted in the liquid regime steadily straighten out with increasing drive until the end of the smectic regime, where the number of half-loops acquires its maximum. Beyond this regime, the pinning influence of the planar defects starts to wane in comparison to the relatively high driving force; hence we observe the number of half-loops monotonically decline with drive. It is worth noting that of the three types of excitations under study, we found the half-loops to be the most resilient structures in the system, with their population being significantly above zero ($\sim 0.2 \dots 0.5$) in the moving-lattice state, even though single kinks (Fig. 6b) and double kinks (Fig. 6c) practically cease to appear in the system well before the onset of this regime. This is consistent with the theoretical findings of Marchetti and Vinokur that for sufficiently large current, half-loop configurations of transverse width smaller than the average separation between the planes become the dominant excitations [44, 45].

Double kinks (Fig. 2c) occur far less frequently (Fig. 6c-/f) than single kinks or half-loops as they require a flux line to assume a spatial structure of relatively higher complexity. The dominant double-kink peak occurs at the end of the liquid regime for orientation A (liquid/smectic for orientation B), at a noticeably higher drive value than that corresponding to the major single-kink peak, which is observed in the center of the liquid(/smectic) regime: As the flux lines evolve from their most distorted configurations in the middle of the liquid regime to straighter shapes at higher drives, the formation of double kinks is facilitated as these vortices start to *loop back* on themselves and reattach to the first defect plane.

These different vortex line excitations mediate thermally activated flux transport with non-linear force-velocity or current-voltage characteristics, as worked out analytically by Marchetti and Vinokur [44, 45] for dilute vortex arrays. The linear transport regime in our system may be characterized by either a rigid flow of flux lines or motion facilitated by double kinks depending on the length of the sample. The non-linear regime is dominated by vortex single kinks and half-loops. There exists a characteristic current scale $J_L \sim 1/L$ that separates the regions of linear and non-linear current-voltage response in the $(L, J)$ plane. Here, $L$ is again the sample thickness in the direction of the magnetic field, and $J$ is the external electric current that exerts a Lorentz driving force $F_d \sim J$ in the direction perpendicular to the defect planes (the $x$ direction); $J_L$ denotes the characteristic current scale, at which flux transport in a sample of thickness $L$ crosses over from a linear to a non-linear regime. Since $J \sim F_d$, it follows that $F_L \sim 1/L$ where $F_L$ is the crossover drive corresponding to the characteristic current $J_L$.

In order to numerically obtain the crossover boundary curve that separates the regions of linear and non-linear transport, we have identified the critical drive strength $F_L$ when the steady-state number of half-loops in the system (with orientation A) starts assuming non-zero values. We repeated this process for 21 systems of varying sample thickness, evenly spaced between $L = 50 b_0$ and $250 b_0$ (Fig. 7a). We then plotted the resulting $F_L$ values against $1/L$ and successfully fit a straight line to the data points (Fig. 7b), thereby confirming the analytical prediction. Note that our best linear fit has a non-zero $y$ intercept 0.04, implying that for infinitely long flux lines ($L \to \infty$) in our system, a non-zero finite drive is required for them to form half-loops. This is likely a consequence of the discretization of the flux lines

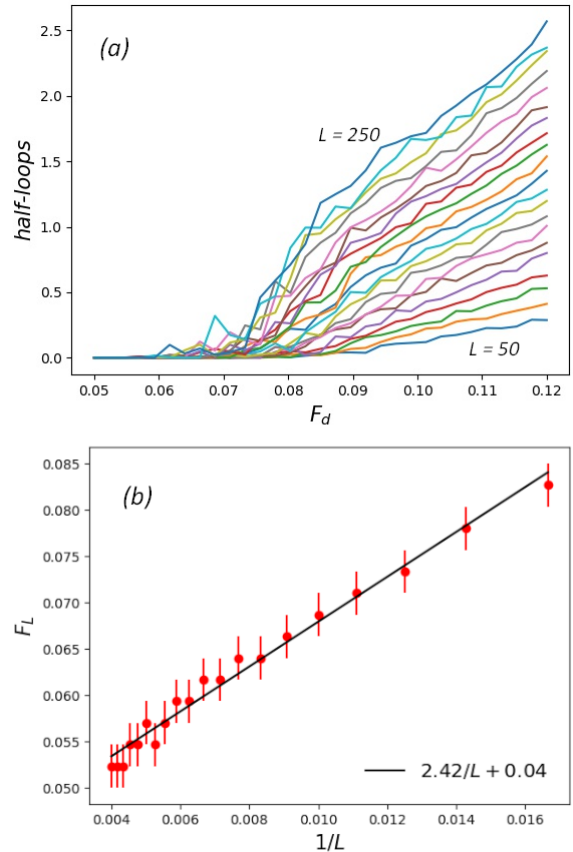

**Fig. 7** (a) Steady-state number of half-loops as a function of drive $F_d$ (units of $\varepsilon_0$) for interacting flux lines in samples with thicknesses varying from $L = 50$ to $L = 250$ (units of $b_0$) in steps of $10 b_0$. (b) Crossover drive strength $F_L$ separating the regions of linear and non-linear response, as a function of inverse sample thickness $1/L$. The onset points in the half-loop curves for different $L$ in (a) are the $F_L$ values used in (b).

along the $z$ direction in our model, which results in a minimum line tension and thus depinning energy necessary to detach a single vortex element.

### 3.3 Widely Spaced Defect Planes

Upon increasing the distance between the planar defects to 50% of the system length, the richness and variety of the observed phenomenology in the depinning regimes encountered for closely placed planes for system orientation A is diminished significantly. The flux lines of all lengths under consideration, except for the longest ($L = 250 b_0$), are then too short to allow for the formation of single-kink (Fig. 8b) or double-kink excitations. For shorter samples ($L < 150 b_0$), the gyration radius valley that marks the partially-ordered regime disappears (Fig. 8a). Consequently, the system is characterized by a flux flow profile resembling samples with columnar defects, *i.e.*, containing a single maximum in the steady-state gyration radius curve that marks the transition of the system directly from the liquid to the smectic regime in the depinning region. For all sample thicknesses considered in this study, the only flux line excitations to appear in any appreciable quantity are half-loops (Fig. 8c), since they require merely one defect plane to form.

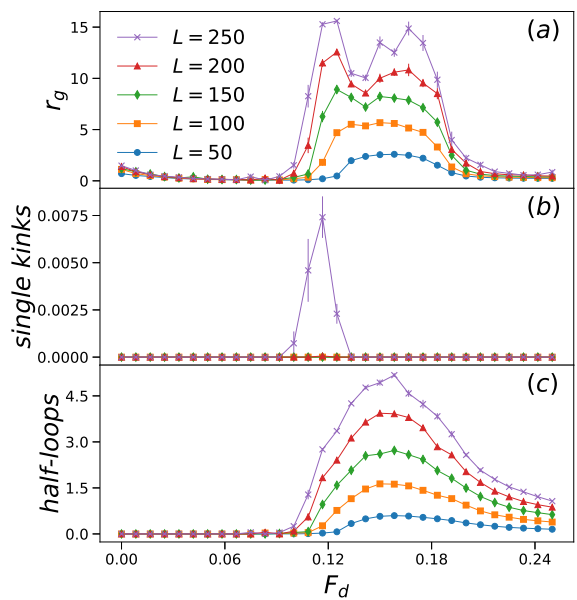

**Fig. 8** Steady-state (a) radius of gyration (units of $b_0$), (b) number of single kinks, and (c) number of half-loops as a function of drive $F_d$ (units of $\varepsilon_0$) for interacting flux lines in the presence of two widely spaced planar defects, in samples of varying sample thickness $L$ (units of $b_0$).

## 4 Conclusion

We have utilized Langevin molecular dynamics simulations to examine a system of driven flux lines in the presence of two planar defects aligned parallel to the magnetic field and perpendicular to the direction of drive. We have probed the steady-state drive dynamics of the system for two horizontal orientations and several sample thicknesses. For closely placed defect planes, we have observed characteristic flux flow regimes that range from a fully pinned stationary configuration at the lowest drive strengths to a perfectly ordered moving lattice at the highest drive strengths. In addition there appear intermediate crossover regimes with vortex matter in different stages of disorder that however manifestly depend on the orientation of the simulation domain. The depinning region in the flux flow profile is broad and displays non-trivial vortex structures. We have characterized these structures by analyzing the unique spatial flux line configurations and excitations that appear during and domi-nate the different drive regimes. These methods supplement measurements of essential global average observables such as local hexatic order and mean gyration radius in the system. The analysis is also aided by rich visualizations of vortices in the various regimes from different perspectives.

The steady-state results and simulation snapshots for the two orthogonal system orientations indicate that strong anisotropic pinning due to parallel planar defects results in a preferred orientation of the vortex lattice that maximizes the number of flux lines encountered by the correlated defects as the former are driven across the sample by an external current. This is evidenced by the tendency of the vortices to arrange into the defect-preferred orientation even when this orientation is orthogonal to the natural lattice orientation for the specific boundary conditions of the system, as is the case for orientation A.

Quantitative measurements of the flux line excitation populations were utilized to detect the boundaries separating distinct crossover regimes for linear and non-linear current-voltage response in the $(L, J)$ or $(L, F_d)$ plane. By identifying the drive strength $F_L$ corresponding to the emergence of half-loops in the system for each sample thickness $L$, we have numerically confirmed that the critical drive strength $F_L$ needed to push the system from a linear to a non-linear transport regime shows a $1/L$ dependence, as analytically predicted by Marchetti and Vinokur [45].

## Author Contribution Statement

All authors participated in conceiving and planning this study, discussing the simulation results, arriving at their physical interpretation, and editing the manuscript. Harshwardhan Chaturvedi and Nathan Galliher performed the numerical computations and carried out the detailed data analysis. Chaturvedi primarily wrote the initial manuscript draft. Ulrich Dobramysl implemented the original simulation code. Michel Pleimling and Uwe C. Täuber provided subject matter expertise and background, and directed the research project.

## Acknowledgments

This material is based upon work supported by the U.S. Department of Energy, Office of Science, Office of Basic Energy Sciences, Division of Materials Sciences and Engineering under Award Number DE-FG02-09ER46613.